\documentclass[lettersize,journal]{IEEEtran}
\usepackage{amsmath,amsfonts}
\usepackage{algorithmic}
\usepackage{algorithm}
\usepackage{array}
\usepackage{textcomp}
\usepackage{stfloats}
\usepackage{url}
\usepackage{verbatim}
\usepackage{graphicx}
\usepackage{cite}
\usepackage{bm}
\usepackage{siunitx}

\usepackage{amssymb}
\usepackage{algorithm}
\usepackage{verbatim}
\usepackage{stfloats}
\usepackage{caption}
\usepackage{subcaption}
\usepackage{multirow}
\usepackage{multicol}
\usepackage{comment}
\usepackage{upgreek}
\usepackage{rotating}
\usepackage{color}
\usepackage{tabularray}
\usepackage{ragged2e}
\usepackage{colortbl}
\usepackage{hhline}
\begin{document}

\title{\textcolor{black}{Superconductor bistable vortex memory for data storage and in-memory computing}}

\author{M. A. Karamuftuoglu, B. Z. Ucpinar, S. Razmkhah, M. Pedram,~\IEEEmembership{IEEE Fellow}
        % <-this % stops a space
\thanks{This work has been funded by the National Science Foundation (NSF) under the project Expedition: (Design and Integration of Superconducting Computation for Ventures beyond Exascale Realization) project with grant number 2124453. (Corresponding author: M. A. Karamuftuoglu)\\
M~A~Karamuftuoglu, B~Z~Ucpinar, S~Razmkhah, and M~Pedram are with the Ming Hsieh Department of Electrical Engineering, University of Southern California, Los Angeles, California, USA (e-mail: razmkhah@usc.edu, pedram@usc.edu)}
}

% \markboth{IEEE TRANSACTIONS ON CIRCUITS AND SYSTEMS—I: REGULAR PAPERS}%

%\IEEEpubid{0000--0000/00\$00.00~\copyright~2024 IEEE}
% Remember, if you use this, you must call \IEEEpubidadjcol in the second

\maketitle

\begin{abstract}

\noindent 
Superconductor electronics (SCE) is a promising complementary and beyond CMOS technology. However, despite its practical benefits, the realization of SCE logic faces a significant challenge due to the absence of dense and scalable nonvolatile memory designs. While various nonvolatile memory technologies, including Non-destructive readout, vortex transitional memory (VTM), and magnetic memory, have been explored, achieving a superconductor random-access memory (RAM) crossbar array remains challenging. This paper introduces a novel, nonvolatile, high-density, and scalable VTM cell design for SCE applications. Our proposed design addresses scaling issues while boasting zero static power consumption characteristics. Our design leverages current summation, enabling analog multiply-accumulate operations —an essential feature for many in-memory computational tasks. We demonstrate the efficacy of our approach with a 32$\times$32 superconductor memory array operating at 20~GHz. This design effectively addresses scaling issues and utilizes current summation that can be used for analog multiply-accumulate operations. Additionally, we showcase the accumulation property of the memory through analog simulations conducted on an 8$\times$8 superconductor crossbar array.
\end{abstract}

\begin{IEEEkeywords}
Vortex memory, crossbar memory array, in-memory computing, superconductor electronics
\end{IEEEkeywords}

\section{Introduction}
\noindent 
Superconductor electronics (SCE) has emerged as a promising beyond-CMOS technology \cite{irds2023}, characterized by its low power consumption and ultra-fast processing capabilities \cite{razmkhahBook}. Recent works have demonstrated the advantages of the SCE across various domains, such as energy-efficient computing \cite{holmes2013energy}, neural networks \cite{Schneider_2022_neuromorphic, Karamuft_neuromorphic}, quantum computing \cite{huang2020superconducting_quantum}, and quantum sensitive detectors \cite{stolz2021superconducting_sensor,natarajan2012superconducting}.

A better approach to utilizing the unique properties of superconductors is using them as a hardware accelerator to complement the well-established CMOS processors. This hybrid approach must include interfaces between CMOS and superconducting logic and transfer large amounts of data across a wide temperature range (e.g., room temperature for conventional CMOS or, at best, 50-70 K range for cryo-CMOS to 4K for superconducting logic). A robust memory is necessary to enable such systems.

Despite the growing demand for fast and efficient superconductor circuits, the lack of a reliable and dense superconducting memory array with high reliability and short access times has remained a challenge. 
Any solution to this problem must enable the storage and retrieval of vast volumes of data to meet the ever-growing computing and memory demands while working at cryogenic temperatures.
%cryocmos memristor optical

To tackle the challenge of cryogenic memory design, researchers explored several design approaches. One proposed solution was utilizing the same Single Flux Quantum (SFQ) technology found in pulse-based SFQ logic to create a memory array. This technology stores data bits by inducing magnetic flux in superconducting loops, relying on Josephson Junction (JJ) switching. Various cryogenic memory designs have adopted this method to achieve promising initial results. The predominant approach within this framework is shift register-based memory design. Mukhanov et al. \cite{mukhanov_93_shiftreg} demonstrated the functionality of a 1024-bit shift register memory with a 19~GHz access frequency.

Similarly, Xu et al. \cite{Xu_2021_rsfq_shiftreg} showcased a shift register memory design featuring bit parallel access, enabling the efficient retrieval of multiple bits of data per clock cycle. However, the shift register structure of the memory restricts random access functionality, a crucial requirement in numerous computing applications. Additionally, the attained memory density remains comparatively low. The other approach uses superconducting non-destructive readout (NDRO) flip-flops as storage elements in a Random Access Memory (RAM) design. Multiple DC-powered superconductor RAM implementations \cite{kirichenko_dcpowered, nagasawa2006dcpowered} utilize transformers for power and data transmission, resulting in a significant layout area. Unfortunately, the area density of such a memory array is relatively low, allowing us to include at most a few kilobits of RAM in a 5$\times$5 mm$^2$ chip. 

Hybrid designs blending CMOS and superconductor circuit characteristics \cite{ghoshal_hybrid}, \cite{duzer_2013_hybrid} offer high density and scalability. However, the access latency increases due to the need to go back and forth between the CMOS and superconducting logic domains, which operate at very different voltage and temperature ranges. Vernik et al. \cite{vernik_2013_mjj} presented a memory design using Magnetic JJs, which is compact and fast. Yet, it is non-scalable due to the high currents required for data writing and the reliability issues with MJJs. Dayton et al. \cite{dayton_magnetic} demonstrated memory implementation with (also magnetic) $\pi$ junctions, providing enhanced reliability and scalability. Nevertheless, including transformers in the design causes bulkiness and vulnerability to external noise.

Nanowires like quantum phase slip junctions (QPSJ) offer the advantage of a small area \cite{Butters_2021_nanowire} and the potential for high-density integration \cite{Zhao_2018_nanowire}. Nevertheless, a major drawback is the absence of a reliable and mature fabrication process for this technology, resulting in low read and write speeds for such memories. Consequently, despite their low power consumption, they fall short compared to cryo-CMOS memory in terms of access time, reliability, and integration density.

Another approach to superconductor memories is using a vortex, which stores bit data in the magnetic field generated by a current flow. Unlike pulse-based approaches, this technology does not necessitate JJ switching and eliminates the need for recovery time after each switch, increasing the speed and power efficiency of the circuits. Vortex-based memory has also shown promising results in terms of its scalability. Tahara et al. \cite{Tahara_1987_vortex_ndro} demonstrated vortex-based NDRO in one of the earliest accounts of this memory. Numata et al. \cite{numata1997vortex} demonstrated a high-density memory application with a relatively lower area. Nagasawa et al. \cite{nagasawa_4kbit_vortex} realized a 4K-bit SFQ hybrid memory using a matrix array of vortex transitional memory (VTM) cells. However, while all the designs mentioned above showed improvement over the speed and integration of conventional SFQ memories, a Vortex-based memory cell requires transformers to manage the data inside. That's why the area usage is still large, resulting in the main drawback of this type of memory structure. Komura et al. \cite{komura2015vortex} introduced a more efficient mutual coupling structure to reduce the area usage of Vortex-based memory. Karamuftuoglu et al. \cite{karamuft_vortex_optimizer_2016} reported a specific optimizer to increase the margin and decrease the area of the Vortex-based memory cell. Semenov et al. \cite{semenov2019very_vortex} demonstrated a large-scale implementation of Vortex-based transitional memory cells. The area consumption was decreased by using self-shunted JJs with a higher critical current fabrication process. However, the mutual inductive coupling in the vortex-based memory made it large and susceptible to magnetic noise. 

This paper presents a nonvolatile Vortex-based memory cell without any transformer. We eliminate the mutual inductive coupling to decrease the memory size and increase its reliability. We demonstrate the correct functionality for both read and write operations. The current design is promising in terms of scalability and non-volatility. We showed a 32$\times$32 matrix array of Vortex-based memory cells operating at 20 GHz. We also demonstrated in-memory computing using this design for an 8$\times$8 matrix array for matrix-vector multiplication at 20 GHz that can be used for fast and efficient artificial neural networks (ANN) design.

The key contributions of this paper are as follows:
\begin{itemize}
\item Eliminating the transformers for reliable operation and higher density  
\item Nonvolatile Vortex memory cell 
\item Memory with a high operation frequency of 50~GHz with no static power consumption
\item Current readout enabling analog in-memory computing on crossbar array
\end{itemize}

\section{Background: Josephson Junctions and SQUIDs}
\noindent 
In normal materials, the phase of the current at different locations is different. However, the phase of the current in the superconductor should be the same since the current is one wave function. If we separate a superconductor into two regions with an insulator, one region will have a quantum mechanical phase of $\phi_1$, while the other will have a phase of $\phi_2$. If the two superconductor regions are in close proximity, their wave function connects and effectively behaves as a single superconductor. In such a scenario, electrical current caused by the phase difference can flow between the two regions without resistance. These currents, known as Josephson currents, flow through this device, which is called Josephson junction (JJ).

JJ is a quantum mechanical device made from two superconducting materials separated by an insulator. JJs can generate SFQ pulses through a quantum phenomenon described by the well-known Josephson equations \cite{razmkhahBook}. A constant phase difference exists between the superconducting wave functions of either side of the junction, resulting in the flow of a supercurrent without any applied external voltage. This behavior of JJ can be characterized by using the DC Josephson equation. In contrast, when there is a constant voltage at the terminal of a JJ, there is an oscillation with the frequency directly connected to the voltage by a constant, described by the AC Josephson equation. 

\begin{figure}[!t]
\centering
    \includegraphics[width=0.8\linewidth]{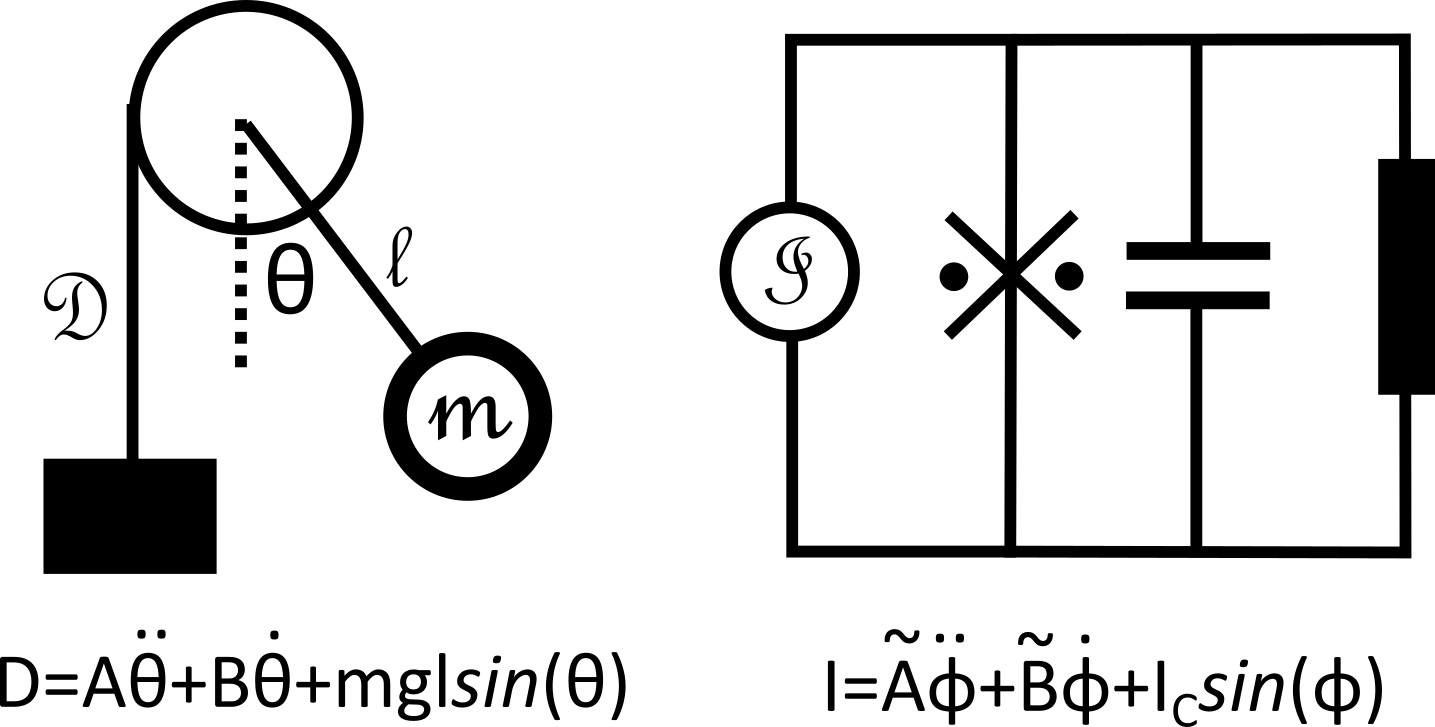}
    \caption{A pendulum with friction and a resistively and capacitively Shunted (Josephson) junction (RCSJ) model are shown with their respective equations. $A$, $B$, $\tilde{A}$, and $\tilde{B}$  are the constant coefficients dependent on the junction's geometry.}
    \label{fig:JJpendulum}
\end{figure}

The Josephson equations exhibit similar behavior to the equations governing the movement of a pendulum. Just as a pendulum returns to its initial state after a change of 2$\pi$ phase, a JJ also cycles back to its original state, generating quantized energy known as Single Flux Quantum (SFQ) as illustrated in Fig.~\ref{fig:JJpendulum}. In SFQ-based circuits, a DC bias is applied to the switching junctions to enhance the switching speed and reduce the required current. This bias current typically falls within 70\%-80\% of a JJ critical current ($I_{c}$). When the combination of an external current and the bias current exceeds the JJs $I_{c}$, the junction becomes resistive, producing a voltage pulse containing a magnetic flux quantum of $\Phi_{0} \approx 2.068 \times 10^{-15}$ Weber. Notice that the Weber unit is equivalent to the Volt-Second and Ampere-Henry units. This phenomenon initiates the generation of an SFQ pulse, after which the JJ reverts to its superconducting state.

Josephson junctions are fundamental active devices in superconducting electronics, much like transistors are in semiconductor electronics. These junctions serve various purposes in electronic circuits, functioning as switching devices, sensors, variable inductors, oscillators (utilizing the AC Josephson effect), and other applications. The Superconductor Quantum Interference Device (SQUID) is a notable circuit utilizing JJs.
A SQUID consists of a closed superconductor loop interrupted by either one JJ, as seen in RF-SQUIDs, or two JJs, in the case of DC-SQUIDs. 

A key characteristic of superconducting loops (rings) is their ability to enclose magnetic flux only in integral multiples of the flux quantum. Leveraging magnetic field quantization, a highly sensitive magnetic detector can be created using a DC SQUID. This device acts as a transducer, converting magnetic flux into voltage, with its sensitivity determined by the magnitude of the flux quantum $\Phi_0$. The output voltage of a SQUID is a periodic function of applied magnetic flux, going through one complete cycle for every quantum flux applied. More precisely, when an external flux is applied to the loop, it generates a screening current to push away the magnetic field until it reaches the quantum of flux ($\Phi_0$). At this threshold, the JJ within the loop switches, allowing a $\Phi _0$ fluxon to be stored inside the loop. In SCE, this mechanism stores fluxons and determines the circuit's state.

When a constant current, known as a bias current, flows through the DC SQUID, it splits equally between the two sides when the SQUID is symmetrical with identical junctions. If the total current does not surpass the critical current of the JJs, which have a lower critical current than the rest of the superconducting loop, a supercurrent will flow through the SQUID.
Now, if a magnetic field is applied to the biased SQUID, it induces a screening current that flows around the loop, opposing the applied magnetic field and effectively canceling the net flux in the loop. The screening current reduces the SQUID's critical current, as the screening current adds to the bias current flowing through one of the Josephson junctions.
As the applied magnetic flux increases, so does the screening current. When the applied flux reaches half a flux quantum, the junctions briefly lose their superconductivity, allowing one magnetic flux quantum to enter the loop before superconductivity is restored. The junctions act as gates for magnetic flux to enter or exit the loop, with the screening current changing direction when the applied flux reaches half of a flux quantum. The screening current decreases as the applied flux approaches one flux quantum, eventually reaching zero when the two fluxes equalize.
Further increments in the applied magnetic flux result in a new screening current flowing in the positive direction, restarting the cycle. This periodic behavior of the screening current concerning the applied flux repeats with a period equal to one flux quantum.

The superconductor-based D-Flip Flop (DFF) is the conventional data storage unit in SCE, featuring a SQUID loop for data storage and a readout circuit. The SQUID's parameters, specifically $L$ and $I_C$, determine the storage capability. If the product of $L \times I_C$ exceeds $\Phi_0$, the SQUID loop can retain the state until a clock pulse triggers the readout circuitry. However, the DFF structure is characterized by a destructive readout process, leading to data loss upon readout.
Non-destructive readout memory (NDRO) structures are implemented to enable multiple readouts without disturbing the internal state of the superconducting loop. However, using NDROs results in bulkier circuits due to the data feedback loop and higher power consumption. Consequently, conventional SFQ-based memory requires rather large inductors within the superconductor loop to store a bit, limiting the memory density necessary for data-intensive computing applications.

\section{Bistable Vortex Memory}
\noindent 
A vortex is a swirling motion of any substance characterized by a rotational flow. In Type-II superconductors, vortices are generated when a magnetic field begins to penetrate the superconducting material in the form of quantized flux. These vortices interact with each other and can exhibit different phases influenced by factors such as the magnetic field, thermal fluctuations, and the pinning effect caused by disorder and defects within the material.
Within the realm of superconducting memory design, a vortex transition refers to the movement or rearrangement of these magnetic vortices within the superconducting material. The presence or absence of vortices and their motion serve as the representations of binary data states. Applying a magnetic field or electrical current makes it possible to write, read, and manipulate the data stored within the memory cell based on the behavior of these vortices.

The VTM cell consists of two superconducting loops and a two-junction interferometer that serves as a sense gate and is magnetically coupled to one of its loops. The cell's operation is controlled by two address signals, $I_x$ and $I_y$, which determine the write and read operations. The control signals must be driven to have positive or negative amplitudes, allowing the data to be encoded with different polarities and values. The cell features a magnetically coupled port for a DC, $I_dc$, which shows that the VTM cell is a volatile memory. Additionally, a sense gate signal, $I_s$, acts as a clock signal, generating the output. In the VTM design, including transformers for input signals can significantly impact functionality, operating speed, and circuit area due to signal propagation times, physical size, and layout constraints. Therefore, in our proposed bistable vortex memory (BVM), we eliminate these design components to simplify the design.

\subsection{Analytical phase-current relations} \label{calcAnalytical}

\begin{figure}[!t]
\centering
    \includegraphics[width=0.85\linewidth]{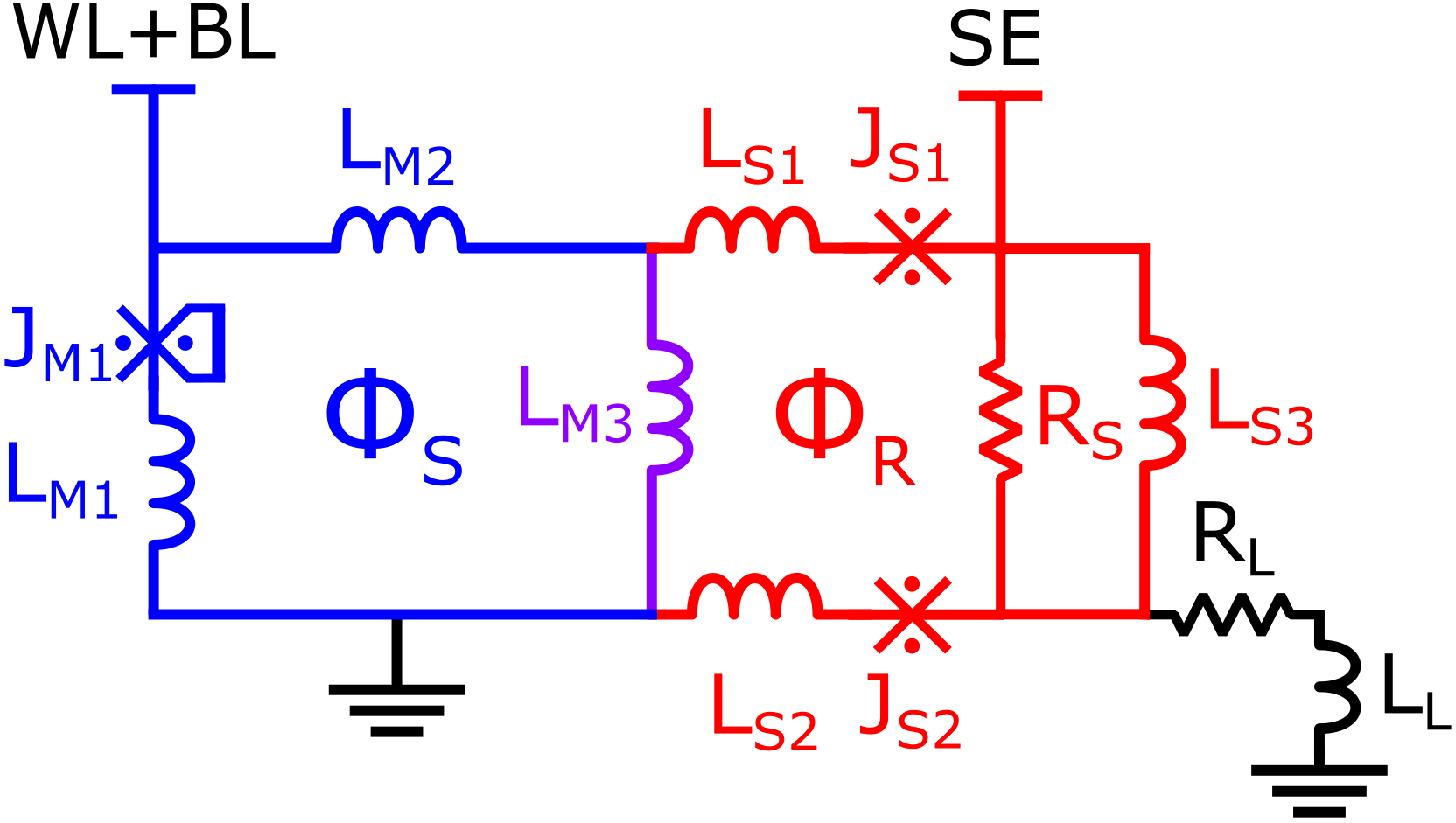}
    \caption{BVM cell model with row and column access lines. Here, the storage loop is drawn in blue, whereas the readout loop is drawn in red. The junction $J_{M1}$ is the only shunted JJ here, and its symbol's vertical line represents the shunted resistor.}
    \label{fig:bvmCalc}
\end{figure}

\noindent 
The design of a BVM cell consists of two connected SQUID loops. The first loop stores the data and the second loop reads the state of the first loop, as seen in Fig.~\ref{fig:bvmCalc}. Depending on the current direction, the cumulative current from the word line (WL) and the bit line (BL) will trap a state of 1 or 0 in the S-Loop. The current from the sense-enable (SE) will cause the readout loop to generate a current at the load element ($L_L$). 

Let's define the direction of the current flowing in the storage loop counterclockwise. Similarly, we define the current flowing in the readout loop clockwise. Further, the direction of bias currents is downward into the superconducting loops. Finally, the direction of the current flow in $R_{S}$ and $R_L$ is to GND. We can then write the current equations for these loops as follows:

\begin{equation}
\small
    \begin{aligned}
        I_{M2} + I_{WL} + I_{BL} = I_{M1}  \\
        I_{M3} = I_{M2} + I_{S1} \\
        I_{S1} + I_{SE} = I_{RS} + I_{S3} \\
        I_{L} = I_{RS} + I_{S3} - I_{S2}
    \end{aligned}
\end{equation}

For the JJs, we can write the DC Josephson equation to relate the JJs phase and current:

\begin{equation}
\small
    \begin{aligned}
        I_{M1} = I^0_{M1} \times sin(\varphi_{M1}) \\
%        I_{M2} = I^0_{M2} \times sin(\varphi_{M2}) \\
        I_{S1} = I^0_{S1} \times sin(\varphi_{S1}) \\ 
        I_{S2} = I^0_{S2} \times sin(\varphi_{S2}) 
    \end{aligned}
\end{equation}
\noindent \ignorespacesafterend where $I^0_{*}$ and $\varphi_{*}$ denote the critical current and phase of the junction $J_{*}$. These equations connect the phase of the SQUID loops to their current. The phase should be a multiple of $2\pi$ for each SQUID loop. Therefore, we can write the phase equations for the storage and readout superconducting loops as follows:

\begin{equation}
\small
\varphi_{M1} + I_{M1}L_{M1} + I_{M3}L_{M3} + I_{M2}L_{M2} = 2M\pi
\end{equation}
\begin{equation}
\small
I_{M3}L_{M3} + I_{S1}L_{S1} + \varphi_{S1} + I_{S3}L_{S3} + \varphi_{S2} + I_{S2}L_{S2} = 2N\pi
\end{equation}

% \begin{equation} \label{eqn:ANPSOvelocityNext}
%  \small
%     \begin{split}
%    	    V^{t+1}_{i} & =	\chi \times (C^{t}_{0} \times V^{t}_{i} \\
%    	    & + C_{1} \times rand(0,1) \times (pbest^{t}_{i} - X^{t}_{i}) \\
%    	    & + C_{2} \times rand(0,1) \times (lbest^{t}_{j} - X^{t}_{i}))
%     \end{split}
% \end{equation}

\noindent \ignorespacesafterend where $M$ and $N$ are positive or negative integer values. When $M = -1$, the internal flux of the SQUID loop, i.e., $\Phi _S = -\Phi _0$, the memory cell's storage state is zero, indicating that the stored value is zero. Conversely, when $M = 1$, i.e., $\Phi _S = \Phi _0$, it represents a storage state of one for the memory cell, with the stored value being one. While a system of nonlinear equations for circuits with more than two JJs does not have a closed-form solution, it can still be solved numerically or approximated \cite{akgun2022pysqif,ilin2019static}. 

The circuit's operation can be divided into four stages: the idle state, the write cycle, the wait cycle, and the read cycle. Initially, all components are idle, establishing the circuit's initial conditions. Data is written into the storage loop during the write cycle, followed by a waiting period for the currents to stabilize. Subsequently, the stored data is read from the sensing loop during the read cycle.

In the idle state, all current sources $I_{WL}$, $I_{BL}$, and $I_{SE}$ are set to zero and $I_{RS}=0$, resulting in $I_{L}$  also being zero. Additionally, the stored data in the storage loop, represented by the variable $M$, is also zero, indicating no stored data is present. Given these conditions (which result in $I_{M1}=I_{M2}$ and $I_{S1} = I_{S2} = I_{S3}$), the equations for the idle state may be written as:
\begin{equation}
\label{eq:initeq2}
\small
    \begin{aligned}
        &\varphi_{M1} + I^0_{M1}\sin(\varphi_{M1})L_{M1} + \\
        &\left(I^0_{M1}\sin(\varphi_{M1}) + I^0_{S1}\sin(\varphi_{S1})\right)L_{M3} + \\
        &I^0_{M1}\sin(\varphi_{M1})L_{M2} = 0
    \end{aligned}
\end{equation}

\begin{equation}
\label{eq:initeq1}
\small
    \begin{aligned}
        &\left(I^0_{M1}\sin(\varphi_{M1}) + I^0_{S1}\sin(\varphi_{S1})\right)L_{M3} + \\ &I^0_{S1}\sin(\varphi_{S1})L_{S1} + 
        \varphi_{S1} + I^0_{S1}\sin(\varphi_{S1})L_{S3} + \\ &\sin^{-1}\left(\frac{I^0_{S1}\sin(\varphi_{S1})}{I^0_{S2}}\right) + I^0_{S1}\sin(\varphi_{S1})L_{S2}  = 0
    \end{aligned}
\end{equation}

By solving Eq.\ref{eq:initeq2} and Eq.\ref{eq:initeq1}, we establish the system's initial state, ensuring that the relevant variables are initialized at their base values. We can determine the initial values of $\varphi_{M1}$ and $\varphi_{S1}$, which are (0,0). Therefore, at the start, both phases are zero, resulting in no current in the respective JJs.

We begin with the initial values established during the idle state. In the write cycle, a value is written to the storage loop by the currents supplied to the circuit from the \( I_{WL} \) and \( I_{BL} \) sources, while \( I_{SE} \) and consequently \( I_{L} \)  remain at zero. The currents \( I_{WL} \) and \( I_{BL} \) are added together for the storage loop. Each of these two currents can be positive, negative, or zero. However, the sum value  \( I_{WL} + I_{BL} \) is what matters.

Initially, \( M \) is zero; however, a successful write operation occurs when \( I_{M1} \) exceeds \( I^0_{M1} \), causing \( J_{M1} \) to switch, setting \( M = \pm 1 \). \( M = \pm 1 \) indicates the presence of a vortex \( \Phi 0 \) in the system. The specific directions of \( I_{WL} + I_{BL} \) determine the direction of the vortex in the loop, resulting in the storage of either logic '0' or '1' values. When the directions of \( I_{WL}\) and \(I_{BL} \) are opposite of each other, the current sum becomes small and cannot change the value in the storage loop.  
The conditions for the write cycle are as follows:
\begin{equation}
\small
\begin{aligned}
    &I_{WL}, I_{BL} \neq 0 \\
    &I_{SE}, I_{L} = 0 \\
    &M = \begin{cases}
        0, & \text{if } I_{M1} \leq I^0_{M1}\sin(\varphi_{M1}) \\
        1, & \text{if } I_{M1} > I^0_{M1}\sin(\varphi_{M1})\\
    \end{cases} \\
    &N = 0
\end{aligned}
\end{equation}
By substituting these conditions, we have the equations for the write cycle,
\begin{equation}
\label{eq:writeeq2}
\small
    \begin{aligned}
        &\varphi_{M1} + I^0_{M1}\sin(\varphi_{M1})L_{M1} + \\
        &\left(I^0_{M1}\sin(\varphi_{M1}) - (I_{WL} + I_{BL})  + I^0_{S1}\sin(\varphi_{S1})\right)L_{M3} \\
        &\left(I^0_{M1}\sin(\varphi_{M1}) - (I_{WL} + I_{BL})\right)L_{M2} = 2\pi
    \end{aligned}
\end{equation}
\begin{equation}\label{eq:writeeq1}
\small
    \begin{aligned}
        &\left(I^0_{M1}\sin(\varphi_{M1}) - (I_{WL} + I_{BL}) + I^0_{S1}\sin(\varphi_{S1})\right)L_{M3} + \\ 
        &I^0_{S1}\sin(\varphi_{S1})L_{S1} + \varphi_{S1} + I^0_{S1}\sin(\varphi_{S1})L_{S3} +\\ &\sin^{-1}\left(\frac{I^0_{S1}\sin(\varphi_{S1})}{I^0_{S2}}\right) + I^0_{S1}\sin(\varphi_{S1})L_{S2}  = 0
    \end{aligned}
\end{equation}

\noindent
Solving Eq.\ref{eq:writeeq2} and Eq.\ref{eq:writeeq1} allows us to explore the relationship between \( I_{WL} + I_{BL} \) and \( I_{M1} \). This analysis is crucial for determining the switching time of \( J_{M1} \). Once we have obtained \( I_{M1} \), we can consequently derive \( I_{M2} \).

During the wait cycle, \( I_{WL}, I_{BL} \) go back to zero, and we utilize the initial values of \( I_{M1} \) and \( I_{M2} \) obtained from the write cycle. This phase ensures a waiting period after writing a value to the storage loop. Consequently, all current sources \( I_{WL} \), \( I_{BL} \), \( I_{SE} \), and \( N \) are set to zero, resulting in \( I_{L} \) being zero as well. However, \( M \) is equal to 1 due to the presence of the stored value. The value of \( M \) will determine the conditions for the wait cycle as follows,
\[
\small
\begin{aligned}
    &I_{WL}, I_{BL}, I_{SE}, I_{L} = 0, \\
    &M = 1,  N = 0
\end{aligned}
\]
By substituting these conditions, we arrive at the equation for the wait cycle. The readout loop equation is the same as Eq.~\ref{eq:initeq1}. The storage loop equation is

\begin{equation}\label{eq:waiteq2}
\small
    \begin{aligned}
        &\varphi_{M1} + I^0_{M1}\sin(\varphi_{M1})L_{M1} + \\
        & \left(I^0_{M1}\sin(\varphi_{M1}) + I^0_{S1}\sin(\varphi_{S1})\right)L_{M3} \\ &I^0_{M1}\sin(\varphi_{M1})L_{M2} = 2\pi
    \end{aligned}
\end{equation}

The stored value induces a change in the current flowing through \( L_{M3} \), influencing the current in the read loop. With the values of \( I_{M1} \) and \( I_{M2} \) obtained from previous equations, we can use Eq.\ref{eq:initeq1} and Eq.\ref{eq:waiteq2} to incorporate these values and determine the initial conditions of \( I_{S1} \) and \( I_{S2} \) for the read cycle.

The stored value within the loop is observed during the read cycle. With \( M = 1 \) showing the stored value '1', resulting in screening current for \( I_{M3} \) in the storage loop, and the initial values caused by it for \( I_{S1} \) and \( I_{S2} \) in the read loop, we proceed to read the stored data. In this case, both \( I_{WL}\) and \( I_{SE} \) are applied to the circuit, leading to a non-zero value of \( I_{L} \). This action triggers the switching of either \( J_{S1} \) or \( J_{S2} \), depending on the value of \( M \), when \( I_{S1} > I^0_{S1} \) or \( I_{S2} > I^0_{S2} \), respectively. The switching of JJs results in the generation of a voltage pulse \( V_{pulse} \), subsequently inducing a current \( I_{L} \) according to the relation \( I_{L} > V_{pulse}/R_L \).
The conditions of the read cycle are as follows:
\[
\small
\begin{aligned}
    &I_{WL}, I_{BL}, I_{SE}, I_{L} \neq 0 \\
    &M = 1, N = 0
\end{aligned}
\]

By substituting these conditions, we have the equations of the read cycle as,
\begin{equation}\label{eq:readeq2}
\small
    \begin{aligned}
        &\varphi_{M1} + I^0_{M1}\sin(\varphi_{M1})L_{M1} + \\ 
        & \left(I^0_{M1}\sin(\varphi_{M1}) - I_{WL} + I^0_{S1}\sin(\varphi_{S1})\right)L_{M3} \\       &\left(I^0_{M1}\sin(\varphi_{M1}) - I_{WL}\right)L_{M2} = 2\pi
    \end{aligned}
\end{equation}
\begin{equation}\label{eq:readeq1}
\small
    \begin{aligned}
        &(I^0_{M1}\sin(\varphi_{M1}) - I_{WL} + I^0_{S1}\sin(\varphi_{S1}))L_{M3} + \\ &I^0_{S1}\sin(\varphi_{S1})L_{S1} + \varphi_{S1} + \left(I_{SE} + I^0_{S1}\sin(\varphi_{S1})\right)L_{S3} +\\
        &\sin^{-1}\left(\frac{I^0_{S1}\sin(\varphi_{S1})}{I^0_{S2}}\right) + \left(I_{SE} + I^0_{S1}\right)\sin(\varphi_{S1})L_{S2}  = 0
    \end{aligned}
\end{equation}

\noindent \ignorespacesafterend The above equations collectively provide the closed-form expressions needed to analyze the BVM cell.

\subsection{Circuit design}
\begin{figure*}[!t]
\centering
    \includegraphics[width=1\linewidth]{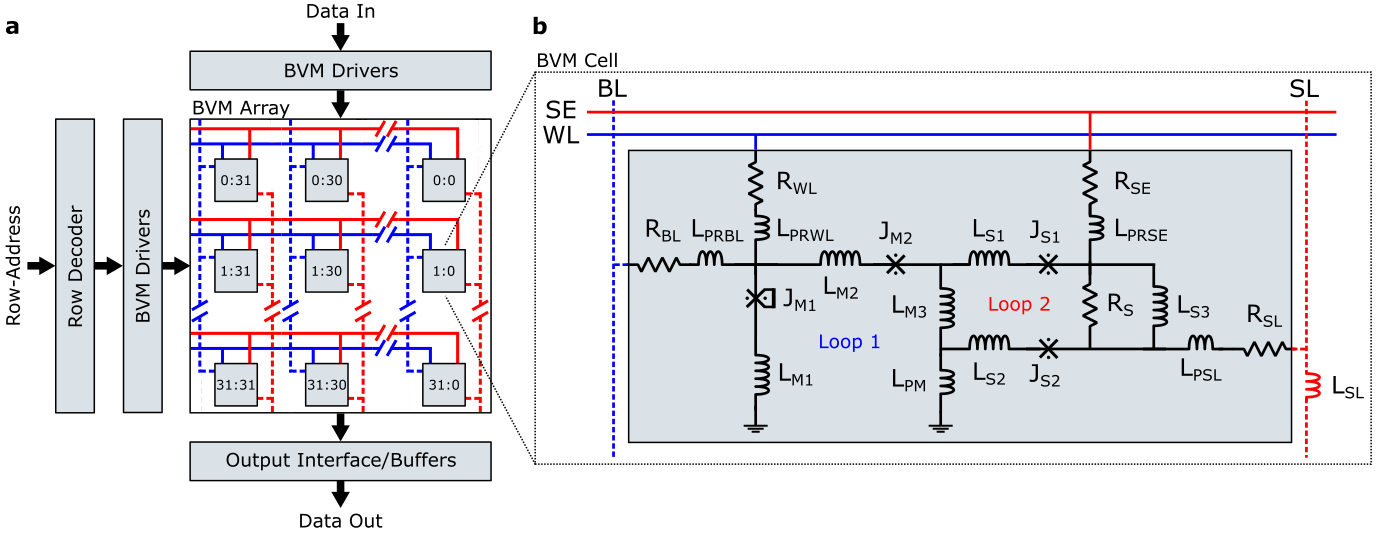}
    \caption{a) The illustration of the BVM cell array with 32 rows and 32 columns. For a larger range of column bits, a column decoder may be necessary to select the correct address. b) the BVM cell's schematic consists of 4 JJs with row and column access lines. The inductances named $L_{P}$ correspond to the parasitics. The Storage Loop stores the data, and the read operation is performed with the help of the Readout Loop. ($R_{WL}$ = $R_{BL}$ = $R_{SE}$ = 20.0 $\Omega$, $R_{S}$ = 3.0 $\Omega$, $R_{SL}$ = 12.0 $\Omega$, $L_{PRWL}$ = $L_{PRBL}$ = $L_{PRSE}$ = $L_{PM}$ = $L_{S1}$ = $L_{S2}$ = $L_{PSL}$ = 0.5pH, $L_{M1}$ = 12.5pH, $L_{M2}$ = 24.5pH, $L_{M3}$ = 8.5pH, $L_{SL}$ = 0.4pH, $J_{M1}$ = $120\mu A$, $J_{M2}$ = $140\mu A$, $J_{S1}$ = $J_{S2}$ = $74\mu A$)}
    \label{fig:bvmSch}
\end{figure*}

\noindent 
BVM cell comprises two superconducting loops, denoted as the S-Loop (or S-Loop for short) and the Readout Loop (or R-Loop for short). Each loop configuration is designed with a pair of JJs and inductance elements, as illustrated in Fig. \ref{fig:bvmSch}. The example of the connectivity of the memory array is given with a size of 32 by 32, providing details about the interconnections and data paths. In the BVM cell, the write and read operations are achieved by the activity of JJs in the loops.

The BVM unit cell stores 1 bit of data; thus, its states are represented by the direction of the circulating current within the S-Loop, either clockwise or counterclockwise. The write and read mechanisms were explained in detail in the previous section. This storage loop comprises $J_{M1}$, $J_{M2}$, and inductances. The stored data on the S-Loop can be read by checking the state of the R-Loop by switching its JJs without changing the stored data. During the read operation, the output current flows through $R_{SL}$, and the results can be observed on the sense line (SL).

One of the main advantages of the BVM is the absence of transformers. A transformer with a good coupling factor is bulky. The WL and the BL control currents are applied over $R_{WL}$ and $R_{BL}$ traces, corresponding to the row and column address signals, respectively. These address signals' values determine the write operation's loop direction. The sense enable (SE) signal provides additional current to the JJs of the R-Loop during the read cycle, which results in switching if the applied and stored currents flow in the same direction.
Moreover, the proposed memory cell is nonvolatile and requires no bias or offset current to maintain the stored data. Therefore, the design does not consume static power in the memory array. The simulations in the following sections are conducted utilizing the SPICE-based Josephson simulator (JoSIM) \cite{delportJoSIM}, and the models of our JJs are based on the SFQ5ee fabrication process developed for SCE at MIT Lincoln Laboratory \cite{MITLL_tolpygo2015}. More details are provided below.

\subsection{Write operation}
\noindent 
In vortex-based memory implementations, a limitation arises from the necessity of Boolean control signals to choose the write operation modes: write '0' and write '1'. This requirement hinders the simultaneous execution of write operations for '0' and '1' within memory arrays. Consequently, the entire write process may need to be completed in two cycles, reducing the write frequency by half.

\begin{figure}[!t]
\centering
    \includegraphics[width=0.96\linewidth]{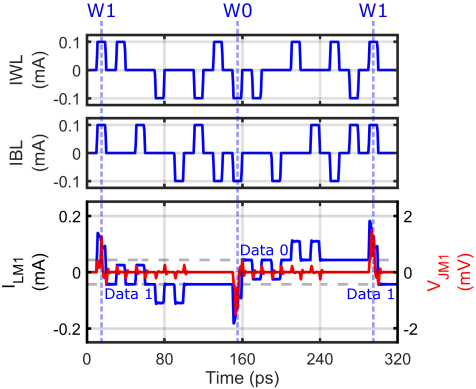}
    \caption{JoSIM result of the BVM cell for the write operation $W$ and with all possible control signal combinations. The simulation is performed at 50 GHz.}
    \label{fig:bvmWriteSim}
\end{figure}

To change the current direction in the S-Loop, write operations are coded into different signal polarities on WL and BL control lines. When the total applied current of $I_{WL}$ and $I_{BL}$ surpasses the critical current of $J_{M1}$, the junction switches, leading to a change in the circulating current's direction. The junction $J_{M2}$ in this design creates a DC-SQUID structure and is used as a non-switching element to prevent feedback from the readout to the storage section. This junction can be removed without a change in functionality (which would save area but also lower parameter critical margins) as explained in the section~\ref{calcAnalytical}. The induced current circulates clockwise if both $I_{WL}$ and $I_{BL}$ amplitudes are positive. Conversely, if they are negative, the current direction changes to counterclockwise. The rest of the address signal combinations are observed as half-selection, and the related operations are shown in Fig.~\ref{fig:bvmWriteSim}.

$I_{WL}$ and $I_{BL}$ govern the operations of the memory cell, as discussed in section\ref{calcAnalytical}. In the analog simulation, the pulse widths of $I_{WL}$ and $I_{BL}$ are configured to 10ps with a 50\% duty cycle, resulting in a 50GHz operating frequency. The initial write operation begins at 10ps, with the change in current flow within the S-Loop being monitored on $L_{M1}$. After the write '1' operation, we test various combinations of address signals corresponding to half-selection cases until 150~ps when the current direction changes due to the negative values of $I_{WL}$ and $I_{BL}$, causing the write '0' operation. The memory state remains unchanged until the subsequent write '1' operation is executed.

\subsection{Read operation}
\noindent 
When the direction of the applied current is aligned with the circulating current in the read loop, an output signal is generated. The $I_{SE}$ polarity indicates the signal polarity in this process. A read-0 operation, where $I_{SE}$ has a negative amplitude, results in an output current when the stored data is '0' and no current when the stored data is '1'. Conversely, a read-1 operation, where $I_{SE}$ has a positive amplitude, generates an output current when the stored data is '1' and no output when the stored data is '0'.
In this context, we used the polarity of stored data 1 for reading. Therefore, the absence of an output current when SE is applied corresponds to the stored data '0'. The level of SE current cannot be higher than \( I_{WL} + I_{BL} \) or it can change the stored data. Maintaining the stored data in the S-Loop is critical to successfully carrying out the non-destructive readout process.

\begin{figure}[!t]
\centering
    \centering
    \includegraphics[width=0.85\linewidth]{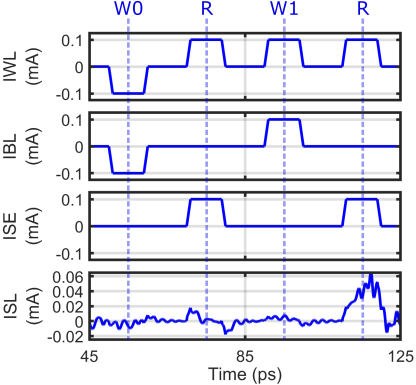}
    \caption{Simulation result of the BVM cell for the read operation \textit{R}. The read is performed for each write operation \textit{W} of '0' and '1', and the output current is observed on the SL. The simulation is performed at 50~GHz, and the load on the sense line is assigned as 12 non-switching junctions with 320~$\mu A$ critical current value each.}
    \label{fig:bvmReadSim}
\end{figure}

Detecting the direction of the current flow in the S-Loop can be achieved by simply triggering the junctions in the R-Loop without altering the data stored in the S-Loop. To achieve this operation, we apply $I_{WL}$ to select the word line and $I_{SE}$ to sense the flowing current. As a result, the sense junctions $J_{S1}$ and $J_{S2}$ continuously switch, thereby generating output current directed towards the $R_{SL}$--$L_{SL}$ path. The readout simulation is shown in Fig. \ref{fig:bvmReadSim}.

We execute a series of write operations during the evaluation, alternating between 0s and 1s. Concurrently, we perform successful read operations between these write cycles at an operating frequency of 50~GHz. As a result, we observe the output current generated across the SL inductance, $L_{SL}$, when the stored data is '1'.

\subsection{Design considerations and trade-offs}
\noindent 
We analyze various inter-dependencies and trade-offs in the BVM cells. The SQUID structures offer flexibility that allows modifications on the critical current ($I_{C}$) of JJs and loop inductance ($L$) to maintain the storage behavior in the S-Loop while satisfying the criterion for flux storage $I_{C} \times L > \Phi _0$. Adjustments may involve increasing the $I_{C}$ of JJs while decreasing $L$, or vice versa. In many series cells (rows or columns), reducing the critical current of BVM's switching junction $J_{M1}$ is the way to lower the required current on WL and BL. This approach helps with the scalability and peripheral circuit design.

Utilizing a single JJ ($J_{M1}$) is adequate for establishing an RF-SQUID storage loop structure for data storage. Adding a non-switching junction $J_{M2}$ enhances stability in our design. Introducing the resistor $R_S$ into the R-Loop in parallel with $L_{S3}$ improves the robustness of the cell, albeit with a slight increase in the cell area.
An alternative modification involves integrating WL and SE to create a unified control signal. This adjustment raises the magnitude of the required current in the control line but reduces the cell area usage. Consequently, during read operations, only the control line WL is activated. In such instances, simultaneous read and write operations can be achieved by activating BL and WL signals. Maintaining these signals separate from each other enhances controllability and distinguishes between read and write operations. %Here, the SE signal acts similarly to the clock signal in the SFQ circuits.

In the BVM design, the inductance $L_{SL}$ on the SL column is crucial for signal propagation to the readout circuit. An alternative approach is to use a stack of non-switching JJs on the SL column to replace the inductance. Since the JJ stack does not require shunt resistance, it occupies a smaller area than $L_{SL}$, thereby reducing the circuit layout area and improving scalability within the BVM array. The time constant for the read operation on each row is determined by the ratio of $\sum_i^{n} L_{SL}^{i}$ to $R_{SL}$,  where index $i$ shows the row ID and $n$ denotes the total row count.
In this case, the inductance introduced by the JJs must be considered. In our design, these stacks of junctions are referred to as $J_{SL}$.

\subsection{Memory array demonstration of 32$\times$32}
\begin{figure}[!t]
\centering
    \includegraphics[width=0.9\linewidth]{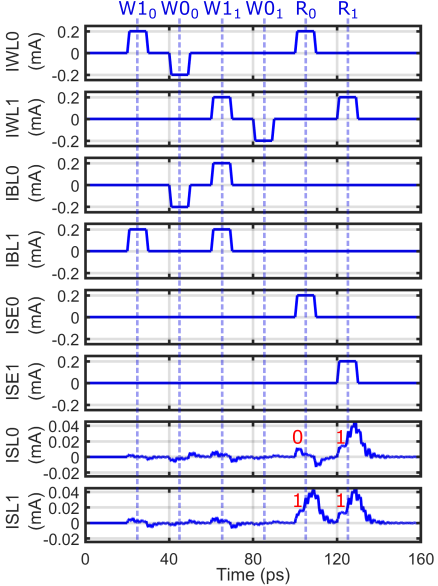}
    \caption{The simulation of the memory array with two rows and two columns. The write operation \textit{W} is completed with two cycles per row, and the read operation is represented with \textit{R}. The subscripts in the operation label correspond to the row index. The stored data in the first and second rows are '01' and '11'. The simulation is conducted at the frequency of 50~GHz, with each sense line having a load of 12 non-switching 320~$\mu A$ junctions.}
    \label{fig:bvm2X2Sim}
\end{figure}

\noindent 
In the memory design featuring BVM cells, distinct write operations are necessary for storing data corresponding to '0' and '1'. To test the functionality and validate the performance of BVMs, a memory configuration of 2$\times$2 has been designed using the same parameter set reported in Fig. \ref{fig:bvmSch}, and the simulation result is shown in Fig. \ref{fig:bvm2X2Sim}. 

At the beginning of the simulation in Fig.~\ref{fig:bvm2X2Sim}, the consecutive write operations are performed for each row, collectively requiring four write cycles at 50~GHz. Upon storing the data in the rows, each row is individually read by enabling WL and SE signals. For each stored value of '1', the output current is observed on each SL, confirming the successful read operation and providing the result.

\begin{figure}[!t]
\centering
    \includegraphics[width=0.85\linewidth]{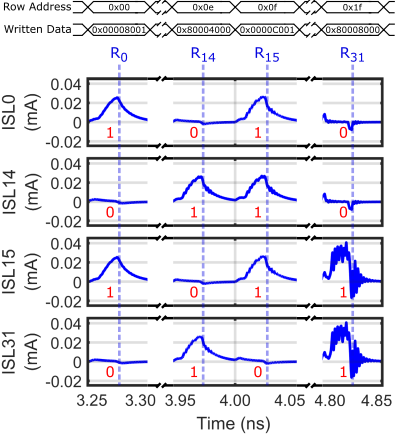}
    \caption{Simulation result for the memory array with 32 rows and 32 columns. $R_0$ corresponds to the read operation of row 0, and $I_{SL0}$ represents the data on column zero. The simulation is performed at 20~GHz, and the load on the sense line is assigned as a non-switching junction with 320~$\mu A$.}
    \label{fig:bvm32x32Sim}
\end{figure}

Even though one of the primary characteristics of the BVM is the ultra-fast access time, that is, how fast information can be written and read, it is not feasible to overlap 10~ps pulses to achieve ultra-high-speed write/read operations. Therefore, the operating frequency was lowered from 50 GHz to 20 GHz for the memory array with 32 rows and 32 columns. The write/read cycle time is stretched for the simulation shown in Fig.~\ref{fig:bvm32x32Sim}. 

The simulation of a memory array with 32 rows and 32 columns operating at 20~GHz is shown in Fig.~\ref{fig:bvm32x32Sim}. In this design, $R_{SL}$, $L_{PSL}$, and $L_{SL}$ are 16~$\Omega$, 0.3~pH, and 0.3~pH, respectively. During the read operation, a current appears on the output path when the stored data is '1'; there will be no current when the stored data is '0'. If the time allowed for dissipating this current is insufficient, a residual current remains on the output path.
Throughout the simulation, the rise/fall time of the input signals is slightly modified to mitigate the current residue between cycles while maintaining a 50\% duty cycle. This approach ensures the integrity of the output data, although it results in a slight drop in the output's current level. Nonetheless, the distinction between '0' and '1' state outputs remains large enough for interface circuits. The operational characteristics and performance are confirmed with the successful completion of the reading process. Although there is some ripple in the output current of rows near the load, the current on the SL surpasses a certain amplitude. Therefore, the resultant current can be detected, and slight differences are ignored at the interface circuit, as explained in the following section. 

\section{Peripheral circuits}

\subsection{Driver}
\noindent 
The peripheral's operating speed can be a bottleneck in high-speed BVM designs. One promising circuit solution is the amplified SFQ to DC converter \cite{razmkhah2021amplifier}, which generates the necessary current at high frequencies. This circuit serves as an interface between pulse-based circuits and BVMs. When an SFQ pulse is received, the circuit converts it to a voltage level, causing the intrinsic state of the SQUID to flip and allowing the circuit to generate current for the output peripherals. Each control signal driving the BVM is separately generated by its dedicated SFQ converter. This configuration enables each control signal to be produced by high-frequency SFQ circuits on the same chip.

To further improve the scalability, a multistage superconductor voltage amplifier can be interfaced with BVM \cite{razmkhah2021amplifier}. The amplifier circuit comprises a series of SQUID loops, wherein the SFQ-to-DC output can be connected to the amplifier input.
A common DC bias is applied over the series of SQUIDs, thereby highlighting the cumulative effect of the series arrangement on the overall voltage output. Moreover, the total voltage observed is the product of the voltage across a single SQUID multiplied by the total number of stages in the series configuration. The driver circuit should support both positive and negative amplitudes at its output to ensure accurate write operation. The differential nature of the SQUID amplifier allows us to put the ground point in the middle and supply both positive and negative inputs from a single circuit.
Consequently, the amplifier circuit enhances the drivability of BVM and increases the number of BVMs that can be integrated into a single row or column.

Given the similarity in control signals between VTM and BVM cells over their respective address ports, the VTM's bipolar signal transmitter can also be utilized as a driver for BVM \cite{karamuft_vortex_optimizer_2016}. This transmitter consists of two identical components, each comprising JJs and resistors, with one being biased by positive current and the other by negative current. As a result, both positive and negative signals can propagate to the BVM cells upon activating either side of the transmitter.

\subsection{Readout}
\begin{figure}[!t]
\centering
\begin{subfigure}{1\linewidth}
    \centering
    \includegraphics[width=1\linewidth]{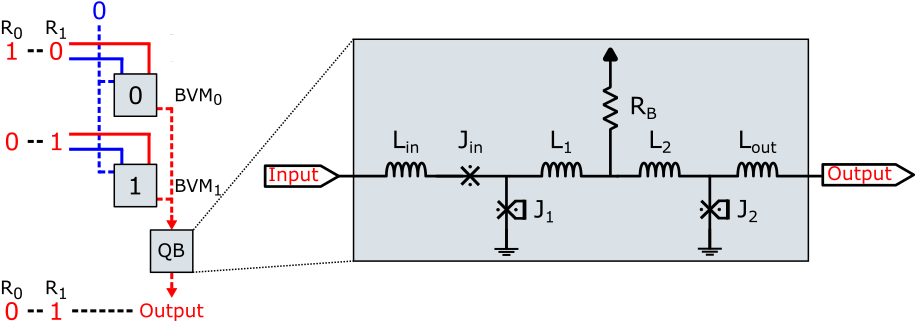}
    \caption{Testbench for the BVM array with the readout QB circuit. The sense line has eight non-switching junctions \textcolor{black}{with 500~$\mu A$ critical current value each.}}
    \label{fig:bvm2X1QB_tb}
\end{subfigure}
\hfill
\vspace{0.5mm}
\begin{subfigure}{1\linewidth}
    \centering
    \includegraphics[width=0.85\linewidth]{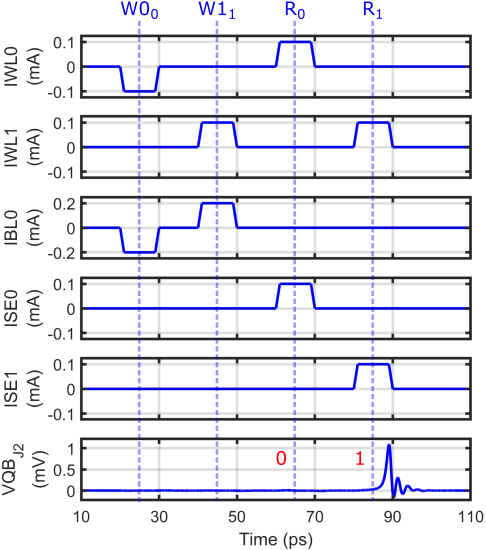}
    \caption{Simulation result at 50~GHz operating frequency.}
    \label{fig:bvm2X1QB}
\end{subfigure}
\caption{Evaluation of the BVM array with a readout QB circuit. Data values 0 and 1 are written into the BVMs in rows 0 and 1, respectively. $W0_{0}$ represents the write-0 operation in row 0, whereas $R_{0}$ corresponds to the read operation in row 0. If the stored value in the BVM cell is 1, the QB cell generates a pulse at the output.}
\label{fig:bvm2X1QBall}
\end{figure}

\noindent 
A peripheral circuit, such as a comparator, is required for analog-to-digital conversion. A readout quantizer buffer (QB) circuit converts the input current into an SFQ pulse in RSFQ technology \cite{razmkhah2023hybrid}. The QB operates similarly to an asynchronous quasi-one junction SQUID circuit \cite{razmkhahBook}. Ensuring the input current meets the SL's required current level is crucial. Once the input current exceeds a predetermined threshold, the circuit generates an SFQ pulse at the output, finalizing the conversion after the read operation on the BVM. The QB circuit is depicted in Fig.~\ref{fig:bvm2X1QBall}.

The QB circuit, which has a simple structure with three JJs, generates an SFQ pulse upon receiving an input current sufficient to excite J1. The QB circuit is modified to detect a single BVM output in the test setup. Two BVMs are placed in a single column, with the sense line connected to a single QB. Data 0 and 1 are initially written to the BVMs in rows 0 and 1, respectively. Subsequently, each row is read sequentially, generating an SFQ pulse on the QB output when the BVM data is 1. This QB circuit successfully establishes the output interface for pulse-based superconductor electronic logic circuits.

In neuromorphic computing, activation functions involve comparing input signals to a specific threshold value to make decisions. These operations are performed by neurons, particularly their soma parts, which are responsible for processing and integrating input signals. Threshold circuits in neuromorphic systems that utilize superconducting components offer promising alternatives for achieving similar functionalities to those found in BVM readout circuits \cite{karamuftHighFaninhybrid, crottyNeuron2010}. To generate digital SFQ outputs, the BVM's SL can be connected to the input of a neuron's threshold gate, generating an SFQ pulse in response.

\section{Analog In-memory computing with BVM crossbar arrays}

\subsection{Structure and operation}
\noindent 
Within the framework of conventional computing systems, performing operations on data involves transferring data to a processing unit and its subsequent return to memory for writing. As a result, the fetch cycle leads to significant latency and energy consumption. For in-memory computing, the operations are carried out within a computational memory unit by leveraging the intrinsic properties of memory devices to eliminate the requirement for data transfers to the processing unit via data buses.

Each memory unit generates an output current on the sense line towards a common ground in the BVM array. Since the SL resistor $R_{SL}$ isolates each memory unit within the same column, there is no interaction among the memory units while the output current is generated. Therefore, by enabling multiple rows with SE signal application, a simultaneous output current generation event occurs, resulting in a summation of multiple row outputs on the sense line.

\begin{figure}[!t]
    \centering
    \includegraphics[width=0.7\linewidth]{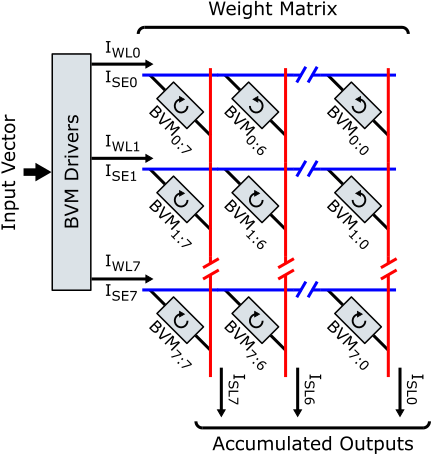}
    \caption{General concept illustration of in-memory computing with the multiply-accumulate operation. The test case incorporates the BVM cell array with eight rows and eight columns. Unlike the general memory purposes, multiple rows are simultaneously read, and as a result, the current accumulation is observed on each SL.}
    \label{fig:bvmInMemSch}
\end{figure}

\begin{figure}[!t]
    \centering
    \includegraphics[width=0.88\linewidth]{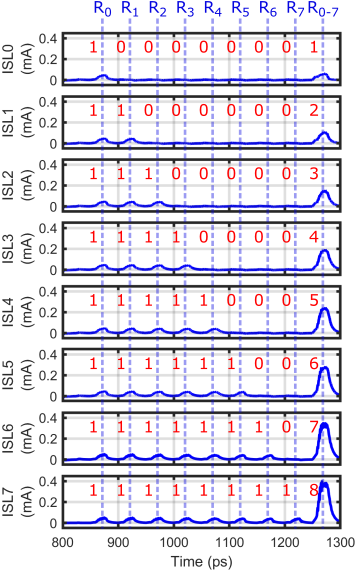}
    \caption{Simulation result of BVM array with enabled multiple rows. $R_0$ corresponds to the read operation of row zero, and $I_{SL0}$ represents the data on column zero. The written numbers represent the output amplitude level. The simulation is performed at 20~GHz, and the load on each sense line is assigned as 12 non-switching junctions with 500~$\mu A$.}
    \label{fig:bvmInMemSim}
\end{figure}

To demonstrate the capabilities of BVM, we designed 8$\times$8 memory as a testbench shown in Fig. \ref{fig:bvmInMemSch}. The current on the load inductance at the end of the sense line is used to dictate the accumulation operation. After writing the data into corresponding rows, the simulation reads individual lines. In the final test case, all rows are simultaneously read, and the current accumulation is observed, confirming the applicability of BVMs for the applications of in-memory computing and neural network computations. The related simulation result showing the read operations is presented in Fig. \ref{fig:bvmInMemSim}. 

Throughout the simulation at 20~GHz operating frequency, the written data across rows zero to seven exhibits a decreasing number of 1s as the row index moves from one to the next. Write-0 operations for all rows can be accomplished in just one cycle due to simultaneous access of multiple rows. The remaining write-1 operations can be completed by accessing one row per cycle. Upon performing each read operation \textit{R} on a single row, a change in the SL current is observed if the stored data is '1'. Subsequently, at the end of the simulation, where all rows are accessed simultaneously for the read operation $R_{0-7}$, different output current levels, depending on the number of stored '1's, are observed in the SL. Such functionality highlights BVM's computing capabilities. In this simulation, the BVM parameters are kept the same values reported in Fig. \ref{fig:bvmSch}.

With its inherent two-state operation, the BVM cell is naturally suited for bitwise operations. In the previous example, each row is assigned to a single input bit, allowing for simple and efficient bitwise operations. The states can be modulated to store multi-bit weights in a neural network and thus enable multi-bit computations. Indeed, multiple rows in the BVM array can be allocated to a single weight. 

\subsection{Applications}
\subsubsection{Multiplication}
\begin{figure}[!t]
    \centering
    \includegraphics[width=1\linewidth]{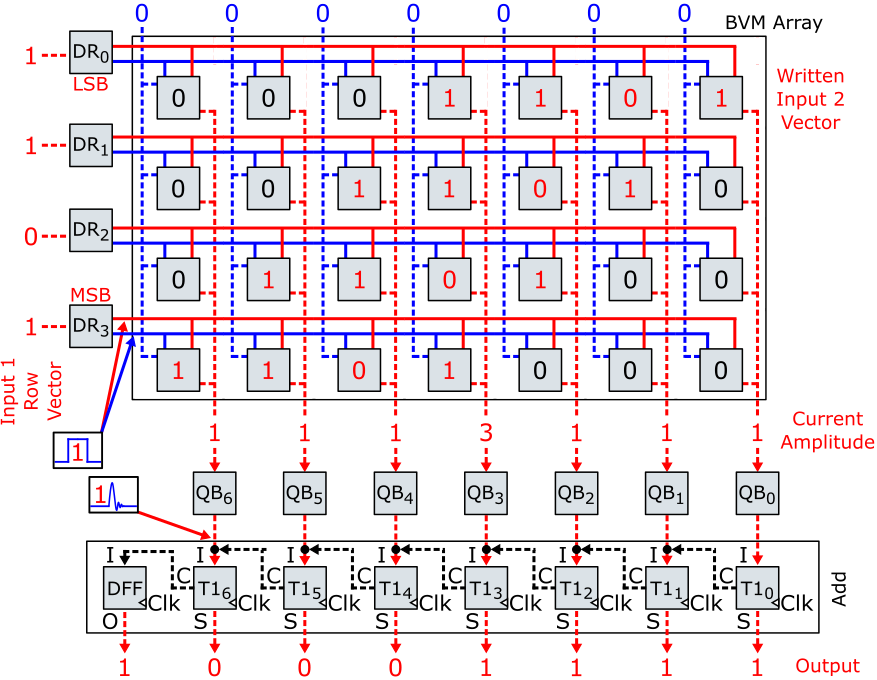}
    \caption{A 4$\times$4 bit multiplier circuit using the BVM array and a carry-shifting circuit. The multiplicand (11$_d$ = 1011$_b$) is the input for the row control signal, while the multiplier (13$_d$ = 1101$_b$) is stored in the BVM array. After receiving a clock signal, the resultant product (143$_d$ = 10001111$_b$) is observed at the output. Here, the T1 cell performs binary addition with a synchronous sum and asynchronous carry output. Input, carry out, sum, and clock signals of the T1 cell correspond to $I$, $C$, $S$, and $CLK$, respectively.}
    \label{fig:bvmMult}
\end{figure}

\noindent 
The BVM array may be used to multiply two multi-bit numbers. As an example, we demonstrate a 4-bit multiplication operation. The first number is written into the 4 BVMs rows so that it occupies column positions $(3,2,1,0)$ in row 1, $(4,3,2,1)$ in row 2, $(5,4,3,2)$ in row 3, and $(6,5,4,3)$ in row 4, respectively. The second number is fed into the read control inputs (SE) of rows 1 through 4, initiating the read operation and enabling column-wise addition. The QB cells placed on each column perform the analog-to-digital conversion and generate one or more SFQ pulses depending on the column's output current amplitude. The output of QBs is the unitary multiplication result of the two numbers fed into a single-cycle carry-shifting circuit. The design is shown in Fig. \ref{fig:bvmMult}.

As seen in Fig. \ref{fig:bvmMult}, the 4$\times$4 bit multiplier circuit is configured to have four rows and seven columns, with the multiplicand and multiplier values set as 11$_d$ (1011$_b$) and 13$_d$ (1101$_b$), respectively. The multiplier number is written into the BVM array, with redundant cells storing data 0. Among the columns, the input of QB$_3$ receives a signal with triple the amplitude of a unit output current, generating three SFQ pulses. At the same time, the rest of the QBs generate one SFQ pulse each. These pulses are then directed into the T1 cells, where internal carry propagation occurs. Once all T1 cells receive their respective clock signals, the final multiplication result 143$_d$ (10001111$_b$) is produced at the output.

\begin{figure}[!t]
\centering
\begin{subfigure}{1\linewidth}
    \centering
    \includegraphics[width=0.40\linewidth]{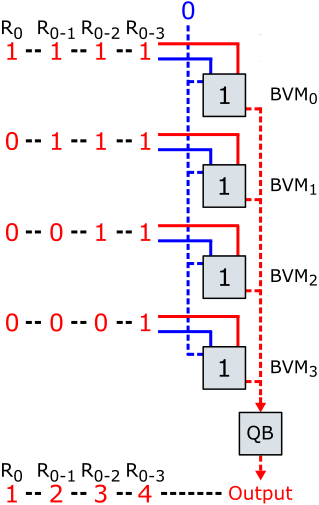}
    \caption{Testbench for the BVM array with a readout QB cell. The column here has four rows. The QB output can be between 0 to 4 pulses depending on the BVM cell values.} %The sense line has eight non-switching junctions with 500~$\mu A$.
    \label{fig:bvm4X1QB_tb}
\end{subfigure}
\hfill
\vspace{0.5mm}
\begin{subfigure}{1\linewidth}
    \centering
    \includegraphics[width=0.85\linewidth]{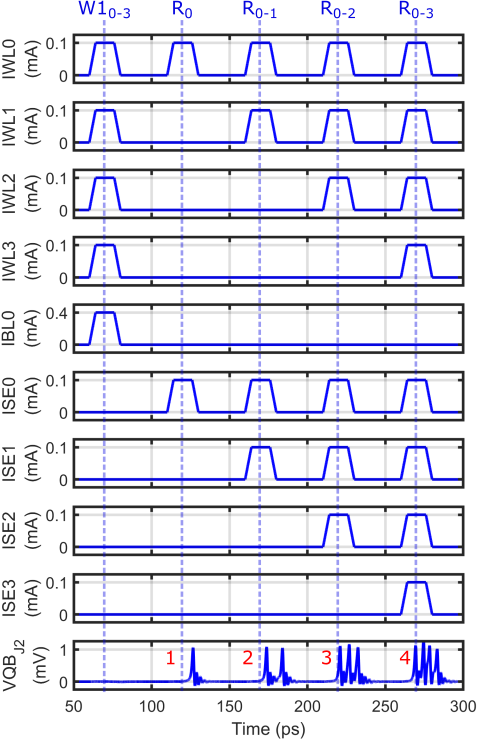}
    \caption{\textcolor{black}{Simulation result.}}
    \label{fig:bvm4X1QB}
\end{subfigure}
\caption{Evaluation of a 4-bit column BVM array with an integrated readout QB. The operations $W1_{0-3}$ and $R_{0-3}$ denote the write-1 and read processes for rows 0 through 3, respectively. In this setup, data value 1 is programmed into all BVM cells. When a BVM cell stores the value 1, the QB cell generates an output pulse. The number of observed pulses at the QB output varies according to the number of rows being read simultaneously.}
\label{fig:bvm4X1QBall}
\end{figure}

To better explain the interface step of the multiplier, we focus on a 4$\times$1 bit BVM array with a readout QB cell, in alignment with the number of rows in the multiplication structure (cf. Fig. \ref{fig:bvm4X1QBall}.) All BVM cells are programmed to store data 1 in this test setup. The current needed to generate an SFQ pulse from the QB circuit is adjusted to the output level of a single BVM cell. As a result, the number of pulses at the QB output reflects the number of 1's stored in the rows that are simultaneously read. The simulation begins with a read operation of a single row and continues until all rows are accessed. This process verifies the intermediate step of the multiplier by generating multiple SFQ pulses corresponding to the accessed BVM cells that store a   1 data value.

\begin{figure}[!t]
\centering
\begin{subfigure}{1\linewidth}
    \centering
    \includegraphics[width=0.95\linewidth]{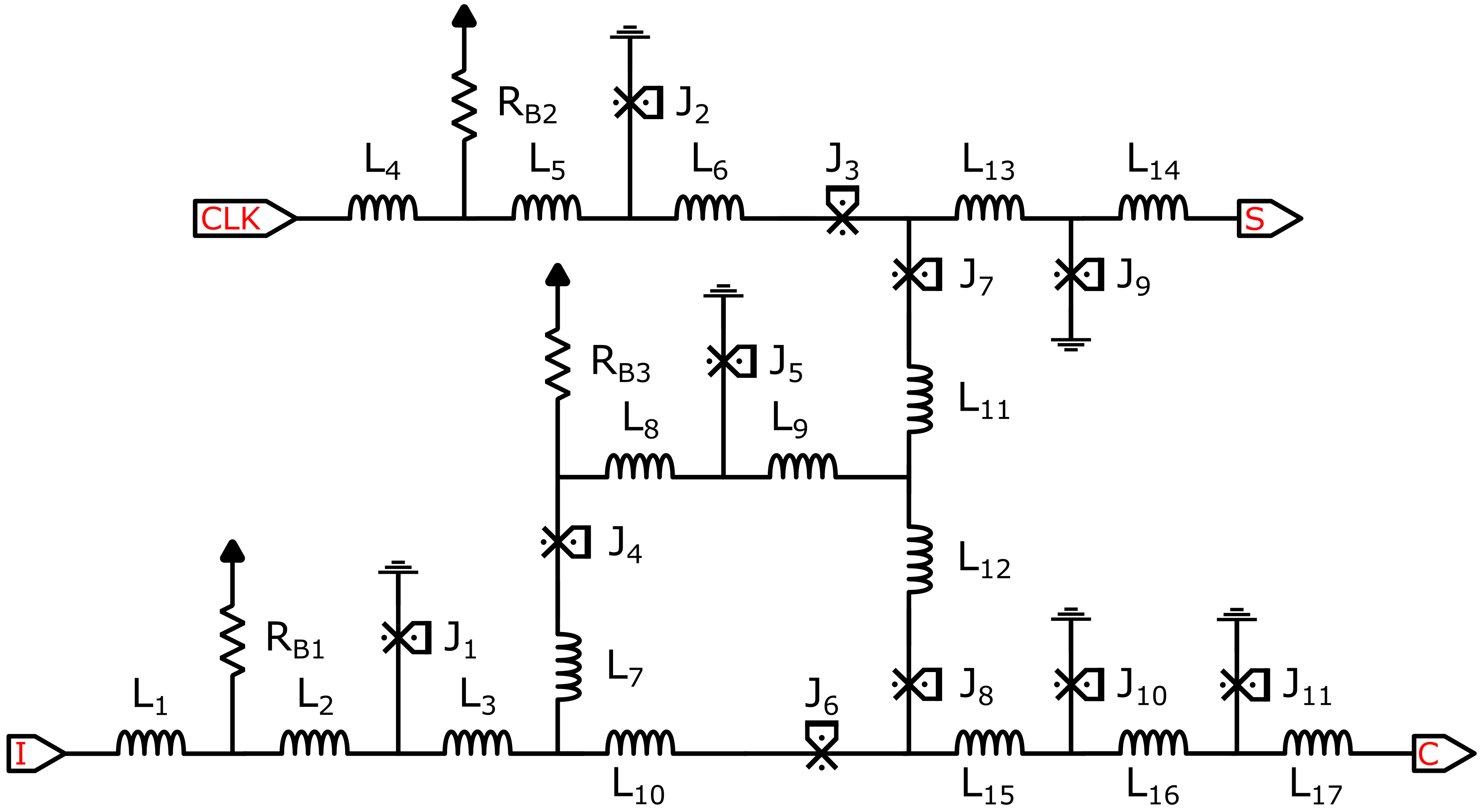}
    \caption{Schematic of T1 cell. ($L_{1}$ = 0.213pH, $L_{2}$ = 1.6pH, $L_{3}$ = 2.028pH, $L_{4}$ = 0.153pH, $L_{5}$ = 0.6pH, $L_{6}$ = 2.337pH, $L_{7}$ = 1.219pH, $L_{8}$ = 1.383pH, $L_{9}$ = 5.366pH, $L_{10}$ = 0.905pH, $L_{11}$ = 0.957pH, $L_{12}$ = 1.219pH, $L_{13}$ = 1.009pH, $L_{14}$ = 4.6pH, $L_{15}$ = 1.297pH, $L_{16}$ = 4.644pH, $L_{17}$ = 2pH,  $R_{B1}$ = $R_{B2}$ = 6.8 $\Omega$, $R_{B3}$ = 16 $\Omega$, $J_{1}$ = $J_{2}$ = $350\mu A$, $J_{3}$ = $180\mu A$, $J_{4}$ = $80.3\mu A$, $J_{5}$ = $77.9\mu A$, $J_{6}$ = $105.1\mu A$, $J_{7}$ = $J_{8}$ = $J_{9}$ = $100\mu A$, $J_{11}$ = $86.9\mu A$, $J_{11}$ = $150\mu A$)}
    \label{fig:bvm4X1QB_tb}
\end{subfigure}
\hfill
\vspace{0.5mm}
\begin{subfigure}{1\linewidth}
    \centering
    \includegraphics[width=0.70\linewidth]{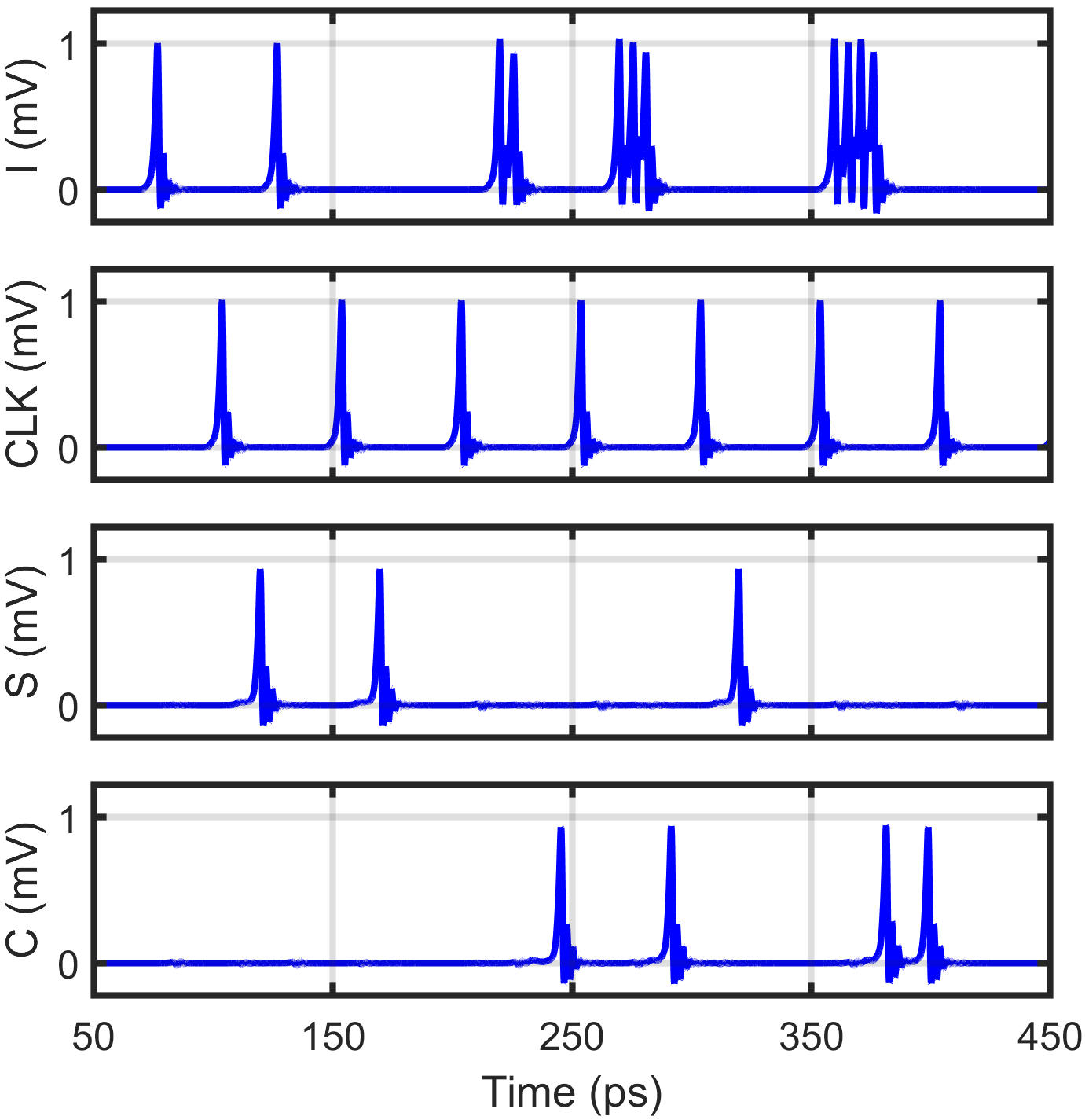}
    \caption{Simulation result. In the plots, pulses are observed from the JJs in Josephson transmission lines (JTLs) \cite{likharev1991a} connected to the T1 cell.}
    \label{fig:bvmT1sim}
\end{subfigure}
\caption{Validating functionality of the T1 cell. Input, clock, sum, and carry-out signals are denoted as $I$, $CLK$, $S$, and $C$, respectively.}
\label{fig:bvmT1all}
\end{figure}

From QBs, pulses propagate into a T1-cell-based adder, and the adder performs the carry-shifting operation, converting the QB outputs to the binary multiplication result. The T1 cell is an SFQ circuit with a synchronous sum and asynchronous carry output \cite{razmkhahBook, bairamkulovT1}. \textcolor{black}{The carry generates an SFQ pulse when two or more input pulses are added, whereas the sum generates an output when an odd number of input pulses are present.} The related circuit with a simulation showing the generation of asynchronous carry and synchronous sum outputs is provided in Fig. \ref{fig:bvmT1all}.

\begin{figure}[!t]
\centering
\begin{subfigure}{1\linewidth}
    \centering
    \includegraphics[width=0.9\linewidth]{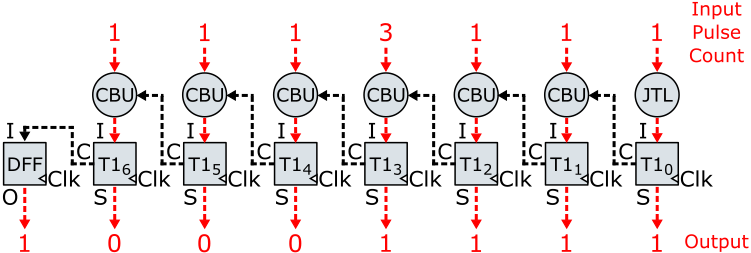}
    \caption{Testbench for the adder part of BVM-based multiplier.}
    \label{fig:bvmAdd_tb}
\end{subfigure}
\hfill
\vspace{0.5mm}
\begin{subfigure}{1\linewidth}
    \centering
    \includegraphics[width=0.85\linewidth]{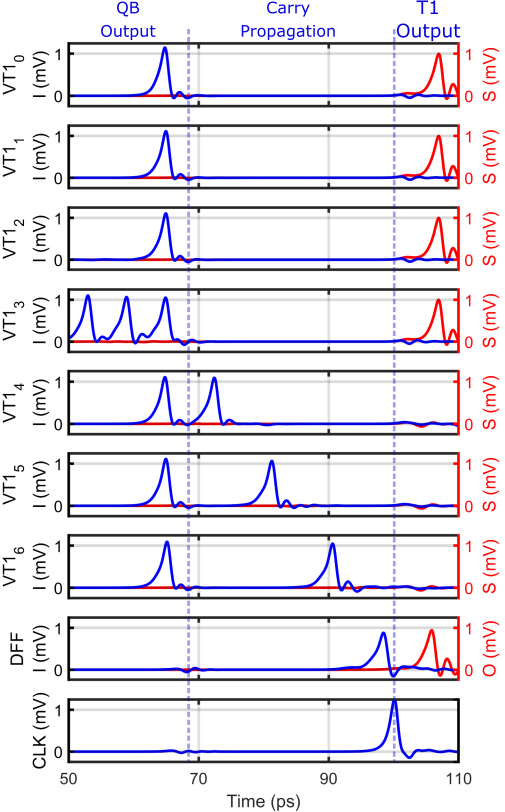}
    \caption{Simulation result for the multiplication of 11$_d$ $\times$ 13$_d$ following the QB outputs shown in Fig. \ref{fig:bvmMult}. The pulses for both the T1 input and output are detected through $J_1$ within T1 and the JJs in the JTLs connected to T1.}
    \label{fig:bvmAddsim}
\end{subfigure}
\caption{Demonstration of the adder used in the multiplier design. The QB outputs can be multiple SFQ pulses on each column. Therefore, a carry-shifting operation is required in the last stage of the multiplier. For the T1, $I$ and $S$ correspond to the input and sum signals, and for the DFF, $I$ and $O$ represent the input and output signals, respectively.}
\label{fig:bvmAddall}
\end{figure}

At the beginning of the simulation (see Fig. \ref{fig:bvmT1sim}), an SFQ pulse is applied to the T1 input, and an output pulse is generated upon receiving the CLK signal. In subsequent steps, multiple pulses are introduced within a short timeframe. \textcolor{black}{A carry output is generated asynchronously for every two input pulses.} An SFQ pulse is observed at the sum output when the pulse count is odd. Due to the asynchronous nature of carry-out generation, multiple carry-out pulses can be observed before the arrival of the CLK signal. This behavior ensures efficient and accurate handling of pulse sequences for the BVM-based multiplier design.

When the input pulses arrive from QBs to T1s, the carry bits are propagated asynchronously from the least significant to the most significant bits. Unlike standard carry propagation methods employed in conventional multiplier structures, this propagation is only performed once, from the least significant to the most significant bits, at the end of the process using a ripple carry T1-based adder. Then, with one clock pulse, the output is generated. To increase the fan-in of the T1 cell, we add a confluence buffer (CBU) into the structure while matching the delay of the first column with the Josephson transmission line (JTL) \cite{likharev1991a}. An example of this configuration is depicted in Fig. \ref{fig:bvmAddall}.

For the example of 11$_d$ $\times$ 13$_d$ multiplication as in Fig.~\ref{fig:bvmAddsim}, all T1s receive an SFQ pulse except the T1$_{3}$ having three SFQ pulses that appear before the 70~ps time point. Since the amplitude of the BVM signal on the QB$_{3}$'s column is larger than the rest of the columns, this QB cell starts generating SFQ pulses earlier than the rest. In this example, the carry propagation occurs between the 70~ps and 100~ps time points. The resultant product (143$_d$ = 10001111$_b$) is observed upon applying a CLK pulse to the cells.

\begin{figure}[!t]
    \centering
    \includegraphics[width=0.88\linewidth]{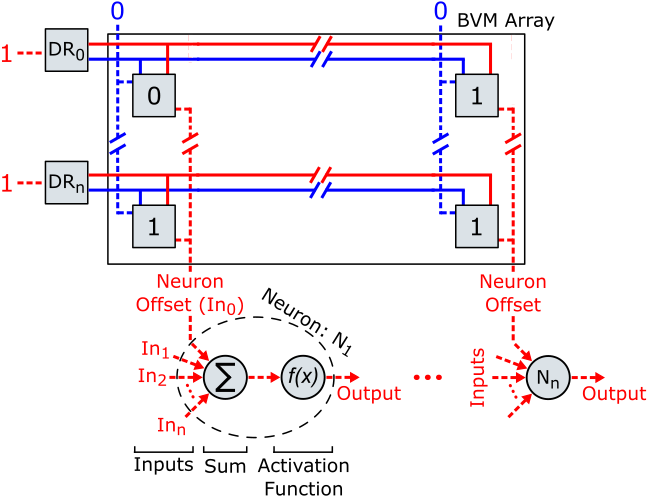}
    \caption{The concept for threshold programmable neurons using BVM array. Each array column can provide a different bias value to its corresponding neuron by storing different data in BVMs. Dedicating specific neuron inputs to apply an offset on the input side alters the required input current to reach the threshold value. This adjustment allows the neuron's behavior to be precisely controlled on-chip.}
    \label{fig:bvmNeuron}
\end{figure}

To compare the multiplier designed using the BVM array with one designed using the conventional RSFQ logic, we have synthesized a 4-bit multiplier using the ColdFlux library and tools \cite{Coldflux2023}. The conventional multiplier circuit uses 13,117 JJs, with an area of 4.14~mm$^2$ and a logic gate depth of 12. The circuit can operate at 44.4 GHz clock frequency, resulting in 270~ps latency and \textcolor{black}{consuming about 1.93~mW of static and 0.05~mW of dynamic power on average}. The proposed design with the BVM design has about ~550 JJs, with an estimated area of 0.236~mm$^2$. The initialization for the multiplier value takes five clock cycles, and it generates the output with one clock cycle for every new multiplicand, operating at 20~GHz and resulting in 50~ps latency. The estimated static power consumption for this design is 0.145~mW, including the peripherals, and consumes 12.23~$\mu$W of dynamic power.

\subsubsection{Threshold-Adjustable Neurons}
\noindent 
Another application of the BVM design is combining this memory with the neuron circuits \cite{karamuftHighFaninhybrid} to achieve an adjustable threshold for these neurons to improve the accuracy of SNN inference \cite{ucpinar2024chip}. 
In neuromorphic computing, a neuron's bias refers to a constant offset or shift in neuron activation. A signal from the BVM circuit can be applied to an appropriately defined input port for a neuron to modify its activation threshold. An example is shown in Fig.~\ref{fig:bvmNeuron}. Here, each column features its dedicated neuron with an adjustable bias. Multiple columns can be assigned to a single neuron to expand its threshold range.

\subsection{Design considerations for BVM-based crossbars}
\noindent 
BVMs for low-power artificial intelligence (AI) applications via in-memory computing create opportunities and pose challenges. Due to fan-in and fan-out limitations on the superconductor hardware, our BVM-based array is much smaller than the standard MAC designs to perform similar operations.

Within the crossbar array, the load on the SL is a crucial factor for each column. Such a design can consist of a single inductor $L_{SL}$ or a stacked set of junctions $J_{SL}$  with a critical current surpassing the maximum level of the achievable accumulated current on the SL. In the case of $J_{SL}$ junctions without an adequate level of critical current, switching events occur during the accumulation, causing an error in the output.
The accumulation operation requires a long enough timeframe to be completed, depending on the number of row values accumulated. 
In other words, the accumulation timeframe must increase linearly with the number of accumulating rows, thus decreasing the operating frequency.

\section{Conclusion}
\noindent 
This paper introduced bistable vortex memory (BVM) as a compact, efficient, and high-performance solution for superconductor electronics. BVM offers zero static power consumption, non-destructive readout, and nonvolatile characteristics. Its simple, transformer-free structure contributes to its compactness and scalability. We demonstrated a 32$\times$32 memory array operating at 20~GHz. Additionally, BVM arrays can perform column addition in a single cycle, making the proposed design an efficient in-memory and near-memory computational unit for matrix-vector multiplication, neural networks, and combinatorial logic implementations. We presented its applications in an efficient multiplier unit and a threshold-adjustable neuronal circuit to emphasize its versatility.

\section*{Acknowledgments}
The authors would like to thank Arda Caliskan (USC) for providing the T1 cell.

\bibliographystyle{IEEEtran}
\bibliography{references}
\end{document}